# Reactor based XPS and other in-situ studies on the Effect of Substitution of Bismuth for Indium in Defect Laden Indium Oxide Hydroxide Bi$_z$In$_{2-z}$O$_{3-x}$(OH)$_y$ on the Photocatalytic Hydrogenation of Carbon Dioxide.

Joel Y.Y Loh[a], Yufeng Ye[a], Nazir P. Kherani [a,b]*

[a] Electrical and Computing Engineering, University of Toronto, Toronto, Ontario M5S 3G4, Canada.
[b] Material Science and Engineering, University of Toronto, Toronto, Ontario M5S 3E4, Canada.
Corresponding author: kherani@ecf.utoronto.ca

**ABSTRACT:** The photocatalytic activity of nanostructured In$_2$O$_{3-x}$(OH)$_y$ for the Reverse Water Gas Shift (RWGS) Reaction CO$_2$ + H$_2$ → CO + H$_2$O can be greatly enhanced by substitution of Bi(III) for In(III) in the lattice of Bi$_z$In$_{2-z}$O$_{3-x}$OH$_y$. This behavior was hypothesized as the effect of the population and location of Bi(III) on the Lewis acidity and Lewis basicity of proximal hydroxide and coordinately unsaturated metal surface sites in Bi$_z$In$_{2-z}$O$_{3-x}$OH$_y$ acting synergistically as a Frustrated Lewis Acid-Base Pair reaction. Reported herein, XPS interrogates this photochemical RWGS reaction transiting from vacuum state to similar conditions in a photocatalytic reactor, under dark and ambient temperature, 150°C, and 150°C with photoillumination. Binding energy shifts were used to correlate the material system's Lewis basicity response to these acidic probe gases. In-situ gas electronic sensitivity and in-situ UV-Vis derived band gap trends confirm the trends shown in the XPS results. The enhanced photo-catalytic reduction rate of CO$_2$ with H$_2$ with low doped 0.05% a.t Bi system is thus associated with an increased gas sensitivity in H$_2$+CO$_2$, a greater increase in the OH shoulder than that of the undoped system under heat and light conditions. The photo-induced expansion of the OH shoulder and the increased positive binding energy shifts shows the important role of photoillumination over that of thermal conditions. The poor catalytic performance of the high doped system can be attributed to a competing H$_2$ reduction of In$^{3+}$. The results provide new insight into how methodical tuning of the Lewis acidity and Lewis basicity of surface Frustrated Lewis Acid Base Pair sites by varying z amount in Bi$_z$In$_2$O$_{3-x}$OH$_y$ enables optimization of the rate of the photochemical RWGS.

## Introduction

Nanostructured indium oxide hydroxide In$_2$O$_{3-x}$OH$_y$ is a photocatalyst for the gas-phase, light-assisted Reverse Water Gas Shift (RWGS) Reaction[1-2], CO$_2$ + H$_2$ → CO + H$_2$O. It has the cubic Bixbyite structure[3] and is synthesized by the partial de-hydroxylation of In(OH)$_3$, where the thermal profile controls the stoichiometry with respect to x and y. At approximately 450-500°C the phase transition from In(OH)$_3$ is complete[4] in order to produce the archetypal Bixbyite end-member In$_2$O$_3$ system. Thermal gravimetric analysis quantifies the de-hydroxylation process, which in its most general form proceeds according to the balanced reaction equation:

2In(OH)$_3$ → In$_2$O$_{2x}$(OH)$_{6-4x}$ + 2×H$_2$O → In$_2$O$_3$ + 3H$_2$O

The OH laden intermediate phase that In$_2$O$_{2x}$(OH)$_{6-4x}$, for the range 0 < x < 3/2, can thus be denoted In$_2$O$_{3-x}$(OH)$_y$. The In$_2$O$_{2-x}$(OH)$_{6-4x}$ thus contains oxide, hydroxide and oxygen vacancies in a cubic Bixbyite lattice, the populations being tailorable through control of the x value[5]. By introducing Bi(III) at different concentrations into the synthesis of the In(OH)$_3$ precursor, it is possible to create partially de-hydroxylated product according to the balanced reaction equation:

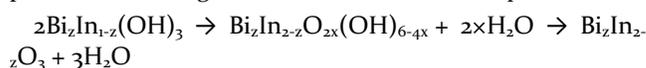

2Bi$_z$In$_{1-z}$(OH)$_3$ → Bi$_z$In$_{2-z}$O$_{2x}$(OH)$_{6-4x}$ + 2×H$_2$O → Bi$_z$In$_{2-z}$O$_3$ + 3H$_2$O

In the aforementioned study, it was discovered that the photocatalytic activity of nanostructured In$_2$O$_{3-x}$(OH)$_y$ for the Reverse Water Gas Shift (RWGS) Reaction CO$_2$ + H$_2$ → CO + H$_2$O can be greatly enhanced by substitution of Bi(III) for In(III) in the lattice of Bi$_z$In$_{2-z}$O$_{3-x}$(OH)$_y$. By systematically varying the atomic fraction of Bi(III) in Bi$_z$In$_{2-z}$O$_{3-x}$(OH)$_y$[6] the CO production rate is 0.67 μmol.g.cat$^{-1}$.h$^{-1}$, 1.12 μmol.g.cat$^{-1}$.h$^{-1}$ and 0.28 μmol.g.cat$^{-1}$.h$^{-1}$ for undoped 0%, low doped 0.05% and high doped 0.5% Bi$_z$In$_{2-z}$O$_{3-x}$(OH)$_y$ respectively. This behavior, which was traced to the strength of the Lewis acidity and Lewis basicity of proximal metal and hydroxide surface sites, is determined by the population and location of Bi(III) sites in $Bi_zIn_{2-z}O_{3-x}(OH)_y$. These surface sites, called Frustrated Lewis Pairs (SFLP)[7-9], play a key role in the gas-phase, light-assisted RWGS reaction catalysed by $Bi_zIn_{2-z}O_{3-x}(OH)_y$. However, since most characterization studies involve samples under air ambient conditions at ambient temperatures or ex-situ characterizations, the actual reactivity of photo-catalytic $Bi_zIn_{2-z}O_{3-x}(OH)_y$ to the various gaseous reactant especially under thermal and photo-illuminated conditions, have not been established.

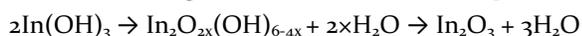

Reported herein, quasi operando reactor XPS measurements, recorded under conditions close to those occurring in a catalytic reactor, provides new insight into how methodically tuning the Lewis acidity and Lewis basicity by varying the substitution level of Bi(III) in the lattice of $Bi_zIn_{2-z}O_{3-x}(OH)_y$, enables systematic optimization of the rate of the photochemical RWGS.

In our quasi-operando XPS experiments, we adopt the following procedure to collect XPS data. The process begins with (i) pre- and post-treatment of $Bi_zIn_{2-z}O_{3-x}(OH)_y$ samples housed in a vacuum tight batch reactor. (ii) Samples are next subject to vacuum, thermochemical and photo-thermal chemical conditions in $H_2$, $CO_2$ and $H_2+CO_2$ atmospheres for 2 hours. Thermal conditions refer to dark and 150°C internal reactor temperature, photo-thermal refers to photoillumination and 150°C internal temperature (iii) The batch reactor is then transferred to a glove box (0.1 ppm $H_2O$ and 0.7 ppm $O_2$) integrated with the XPS spectrometer followed by collection of XPS data. The XPS spectra were all calibrated with the adventitious carbon peak set to a binding energy value of 284.8 eV. In addition, in-situ electrical resistivity as a function of gas/Argon flow ratio was collected in order to determine the gas adsorptive trends of the material system. In order to determine if the quasi in-situ XPS are equivalent to an in-situ measurement, UV-Vis derived band gap measurements[10] under in-situ gas ambient temperature and thermal 150°C conditions were conducted so that the XPS valence band edge derived values approximate that of the in-situ optical spectroscopy derived values. For ease of readability and simplicity, we refer the 0%, 0.05% and 0.5% a.t $Bi_zIn_{2-z}O_{3-x}OH_y$ material system as the undoped $In_{2-z}O_{3-x}OH_y$, low doped and high doped $Bi_zIn_{2-z}O_{3-x}OH_y$.

**Results and Discussion**

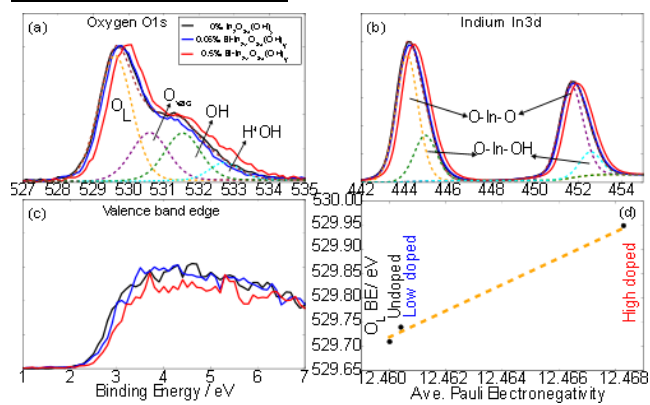

**Figure 1.** (a) O 1s XPS of nanostructured $Bi_zIn_{2-z}O_{3-x}OH_y$, (b) In 3d XPS of $Bi_zIn_{2-z}O_{3-x}OH_y$ and their precursors and (c) valence band edge of $Bi_zIn_{2-z}O_{3-x}OH_y$ and their precursors. (d) the O 1s lattice oxygen binding energy is linear with the averaged metallic ion Pauli electronegativity.

In the case of undoped, nanostructured $In_2O_{3-x}(OH)_y$ and the substituted $Bi_zIn_{2-z}O_{3-x}(OH)_y$ analogues, their O 1s XPS (**Figure 1**) can be resolved into four peaks, whereby these deconvolutions do not exceed the typical maximum linewidths of ~1.4 eV for oxygen species, and hence assigned in increasing energy to oxide lattice $O_L$ (529-530eV), oxygen vacancy $O_{vac}$ (~532.5eV) and hydroxide OH (~533.3eV) and protonated hydroxide groups $H^+OH$ (533.6eV) that arises from sideways coordination of a HOH site on the oxygen lattice, which typically arises from ambient moisture adsorption on the indium oxide surface[11] and is stable even under vacuum conditions. The effective positive charge of a proton bonded to the oxide site of the hydroxide causes the O 1s peak to shift to higher energy than the O 1s of the OH specie thus creating an additional type of OH specie. Since these $H^+OH$ groups are equivalent to a $H^+$ adatom adsorption on a OH group, the $H^+OH$ specie can be used as a way of determining the reactivity of $H_2$ dissociation on the OH group. Additionally, the existence of an oxygen vacancy in the oxide coordination sphere of In(III) or Bi(III) enhances the strength of binding of the remaining oxides, which manifest as a specie of the O1s peak at ~532.4eV. Therefore, the oxide and hydroxide O 1s peaks in these nanostructures, straddle the oxygen vacancy to lower and higher energy, respectively.

Of particular interest in these XPS is the monotonic shift to higher energy of the oxide, oxygen vacancy and hydroxide O 1s peaks with increased level of Bi(III) substitution in $Bi_zIn_{2-z}O_{3-x}OH_y$, which seems to have its origin in the larger electronegativity of Bi(III) relative to In(III). Recall the 6s2 lone pair on Bi(III) is known to be stereo-chemically active, and therefore its spatial direction and extent in the crystal lattice will determine whether the enhanced electron density of the lone pair counterbalances the larger electronegativity of Bi(III) in determining O 1s XPS energies. This also implies that the Pauli electronegativity of the Indium Ion (12.46) is increased with Bi (III) ion (14.14) substitution, and using Vegard's approximation we can show that the binding energy of the oxide subpeak is linearly correlated with the averaged metallic electronegativity (**Figure 1d**).

The XPS and valence band edge results for undoped nanostructured $In_2O_{3-x}(OH)_y$ subjected to vacuum, ambient $H_2$, thermal treatment, $H_2$ at 150°C, and photo-thermal $H_2$ at 150°C under a xenon/mercury-arc lamp are in **Figure 2**. On exposure of undoped $In_2O_{3-x}(OH)_y$ to $H_2$ under ambient conditions, there is a notable increase in the population of hydroxide groups, as seen by the dramatic increase in the O 1s peak around 533 eV. This observation implies the dissociation of $H_2$ on $In_2O_{3-x}(OH)_y$. The concomitant blue shift in the oxide peak around 530 eV likely stems from the charge attraction of the proton on the so-formed hydroxide group and homolytic dissociation of $H_2$ on oxygen lattice sites. The concomitant blue shift seen in the In 3d XPS is consistent with this $H_2$ dissociation scheme as reflected also in the blue shift of the valance band edge corresponding to stabilization of the oxide-like valence band. Interestingly, while OH shoulder expansion under dark and ambient temperature atmospheres of $H_2$ only is seen, thermal conditions reversed the OH shoulder expansion, while maintaining the positive $O_L$ shift which implies that the OH groups due to dissociated $H_2$ on lattice oxide groups are thermally unstable. The pairing of photo and thermal conditions induces far greater $H_2$ dissociation on the OH groups to form new $H^+OH$ groups, and the OH peak intensity increases by nearly 2 times due to homolytic dissociation on oxide lattice to create OH sites. We note that $H_2$ homolytic dissociation on the oxygen lattice sites under thermal and photo-thermal condition are relatively similar since the binding energy shifts of the $O_L$ peak are the same for photo-thermal and thermal only conditions, therefore the main difference between photo-thermal and thermal is the photo-induced increased intensities of the $H^+OH$ specie. This is supported by the valence band results (**Figure 3c**) where additional tailing of the band edge is seen for photo-thermal condition, indicating that these $H^+OH$ states near the valence edge have a higher energy state, ie. these states are deeper into the mid-gap band. We note that the intensities of the In3d peaks in these reactor conditions are normalized with that of the In3d peaks in the vacuum condition, and those normalization factors are applied to the O1s spectrum, which thus indicates that the $O_L$ peaks presented here experiences little signal intensity changes over the transition to these reactor conditions. This additionally implies not all of the generated OH groups are created solely by $H_2$ dissociation on the oxygen lattice sites, however, since the $O_{vac}$ subpeak increases in peak area from vacuum to $H_2$, there is indeed an increase in oxygen vacancy sites. We thus suggest that the increase in OH groups arises in some part from chemisorbed oxygen $O^-$ ions on the surface from atmospheric interaction during synthesis acting as excessively coordinated oxygen-indium surface sites that dissociate $H_2$ to form new OH sites while maintaining the overall net oxidation within the oxygen lattice.

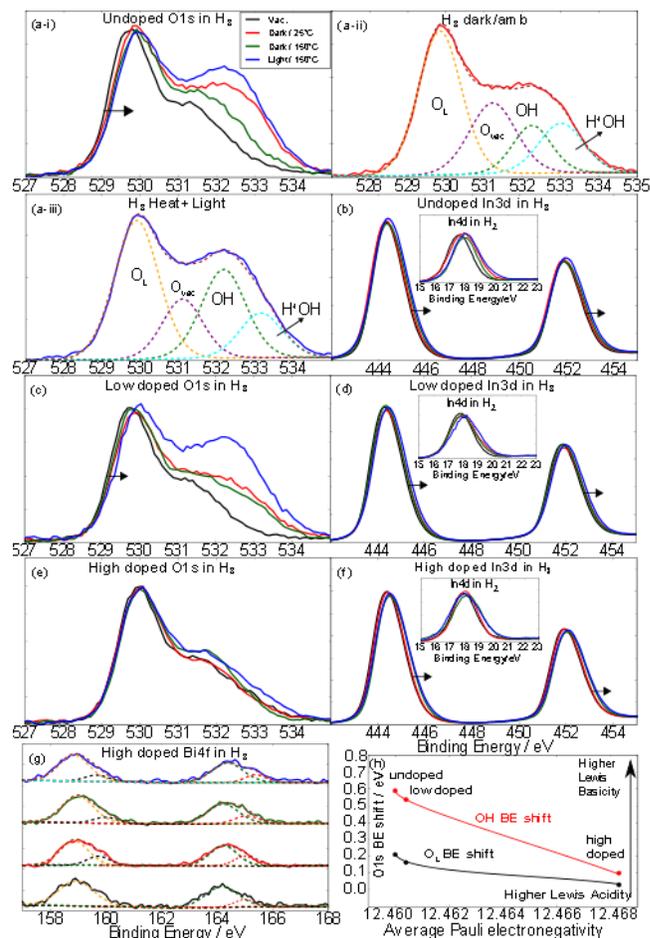

**Figure 2.** XPS of nanostructured $In_2O_{3-x}OH_y$: (a-i) O 1s, (a-i (b) In 3d, of low doped 0.05% a.t $Bi_zIn_{2-z}O_{3-x}OH_y$: (c) O 1s, (d) In 3d, of high doped 0.5% a.t $Bi_zIn_{2-z}O_{3-x}OH_y$: (e) O 1s, (f) In 3d, (g) Bi 4f under vacuum (black), dark ambient $H_2$ (red), thermal treatment, $H_2+CO_2$ at 150°C (green), photo-illumination, $H_2+CO_2$ at 150°C(blue). (h) Value of O and OH binding energy shifts from vacuum to $H_2$ heat and light as a function of the increasing metallic ion electronegativity.

For the low doped $Bi_zIn_{2-z}O_{3-x}(OH)_y$ (**Figure 2c**), the OH shoulder experiences the same large expansion than that of undoped $In_2O_{3-x}(OH)_y$ under photo-thermal condition (**Figure 5b inset**), while binding energy position for the $O_L$ sub-peak is slightly larger than that of undoped $O_L$ subpeak. Under thermal conditions, the wide OH shoulder indicates that $H^+OH$ groups are more thermally stable in low doped $Bi_zIn_{2-z}O_{3-x}(OH)_y$. This hints at the importance of a catalyst not only being able to adsorb and dissociate reactant gases but also having these states stabilized under high temperatures in order to ensure a high kinetic reaction rate. Under photo-thermal condition, the O1s spectrum of the low doped $Bi_zIn_{2-z}O_{3-x}(OH)_y$ are fairly similar in shape to that of the undoped sample (**SI Figure 3**) except for a higher binding energy of the $O_L$ peak for the low doped sample. Since the OH shoulder of the undoped and low doped do not have any binding energy differences, this implies that bismuth substitution did not change the OH group reactivity with $H_2$. For the high doped $Bi_zIn_{2-z}O_{3-x}(OH)_y$ (**Figure 2e**), the introduction of $H_2$ does not induce significant changes in the $O_{1s}$ spectra excepting a small shift in $O_L$ and OH shoulder binding energy under either thermal or photothermal condition. When the $In_{4d}$ spectra is examined between the high doped and undoped/low doped sample, a small tail associated with metallic indium at ~16.5eV is present, indicating that a small amount of indium is being reduced. The $Bi_{4f}$ spectra between the vacuum and photo/thermal conditions generally show a neutral or positive binding energy shift in $H_2$, indicating that bismuth is not being reduced.

Since binding energy shifts can be associated with its protonic affinity or Lewis basicity[12-14], it can be shown that the $O_{1s}$ binding energy of the $O_L$ and OH specie changes from vacuum to photothermal $H_2$ as a function of the averaged electronegativity by following a bowed Vegard approximation (**Figure 2h**). This indicates that the Lewis basicity of the oxygen lattice sites decreases with increasing doping and the OH binding energy shifts in $H_2$ show that the OH groups of low and high doped $Bi_zIn_{2-z}O_{3-x}(OH)_y$ show a much lower Lewis basicity degree such that the low and high doped $Bi_zIn_{2-z}O_{3-x}(OH)_y$ exhibits an increased Lewis acidic behavior under $H_2$ than that of undoped $In_2O_{3-x}(OH)_y$. In all, moderate bismuth substitution of indium in $Bi_zIn_{2-z}O_{3-x}(OH)_y$ exhibits greater Lewis acidity and $H_2$ dissociation than undoped and high doped $Bi_zIn_{2-z}O_{3-x}(OH)_y$.

To support the XPS results, **Figure 3 a-b** shows the electrical conductivity under reactor conditions changes under increasing $H_2$/Argon flow ratios. The slope indicates the adsorption sensitivity[15-16] of the $H_2$ gas by the $Bi_zIn_{2-z}O_{3-x}(OH)_y$ surface. The undoped $In_2O_{3-x}(OH)_y$ shows the greatest increase in conductivity under photo-illumination, with an increased gas ratio threshold before current saturation, which is correlated with the expansion of the OH shoulder seen in XPS. The reversed trends are seen for the high doped $Bi_zIn_{2-z}O_{3-x}(OH)_y$, with conductivity decreasing with increasing $H_2$/Ar flow concentrations, which can be associated with $H_2$ reduction of the indium ion in $Bi_zIn_{2-z}O_{3-x}(OH)_y$ to induce metallic scattering of carriers. **Figure 3d** shows the band gap estimation derived from in-situ UV-Vis spectroscopy. The trends shown in the optical band gap estimation via in-situ spectroscopy and valence band edge fitting generally agree, band gap widening after $H_2$ thermal treatment. This increases our confidence that the results of the quasi in-situ reactor and in-situ UV-Vis spectroscopy are equivalent.

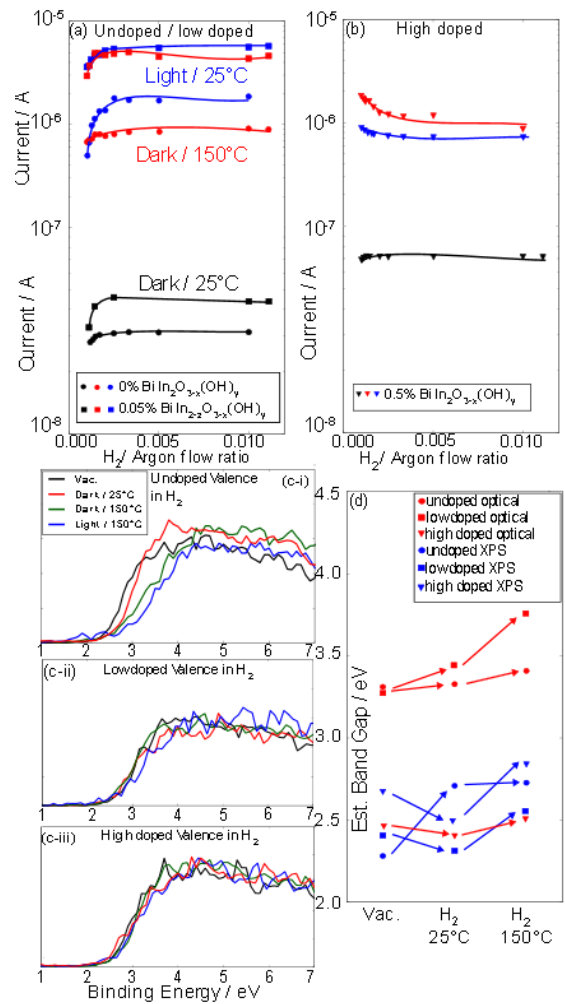

**Figure 3.** Electrical-Gas sensitivity and optical band gap measurement. (a) Undoped and low doped, (b) high doped $Bi_zIn_{2-z}O_3$-$xOH_y$. Circle, square and triangle markers represent undoped, low doped and high doped samples respectively. (c-i-iii) Valence band edge for undoped, low doped and high doped samples respectively. The band gaps are measured by the intercept of the tail of the band edge slope with the binding energy axis. UV-Vis band gap measurement in red with fermi energies determined from valence band edge spectra in blue for undoped (circle), low doped (square) and high doped (triangle) $Bi_zIn_{2-z}O_{3-x}(OH)_y$ in vacuum, $H_2$ ambient temperature and $H_2$ 150°C. Trend lines are visual guides.

We turn our attention to $CO_2$, where **Figure 4** displays the C 1s and O 1s XPS of undoped $In_2O_{3-x}(OH)_y$, low doped and high doped $Bi_zIn_{2-z}O_{3-x}(OH)_y$ under vacuum (black), ambient $CO_2$ (red), photo-illumination, $CO_2$ at 150°C (blue), thermal treatment, $CO_2$ at 150°C (green). Because the presence of carbonates in the $O_{1s}$ spectra is shown in the same binding energy values as that of the OH shoulder, any changes to the OH shoulder of $O_{1s}$ spectra should be associated with surface carbonate formation. The growth of the OH shoulder seen in the O 1s XPS of $In_2O_{3-x}(OH)_y$ on exposure to $CO_2$ under thermal conditions diminishes in 0.05% $Bi_zIn_{2-z}O_{3-x}(OH)_y$ and mostly under high doped $Bi_zIn_{2-z}O_{3-x}(OH)_y$. Contrary to the $H_2$ only condition, photo-illumination under thermal treatment seems to cause these C-O surface species to decay and desorb from the $Bi_zIn_{2-z}O_{3-x}(OH)_y$ surface, whereas the greatest changes to the C1s spectra happens in dark 150°C. The reactivity of the hydroxyl groups in high Bi(III) substituted samples towards $CO_2$ are completely different to the undoped and low Bi(III) substituted samples. For the undoped sample, carbonate formation is seen in dark/ambient temperature, thermal and photo-thermal conditions, where mostly C-O species are mostly seen and less so for O=C-O species, which indicates mostly bidentate, tridentate carbonate formation, some bicarbonate formation and to a smaller extent monodentate carbonate formation. Highly basic adsorption sites favor monodentate formation and less basic sites favor bidentate, bicarbonate formation, hence the C1s spectra supports the more Lewis basic nature of the undoped $In_2O_{3-x}(OH)_y$. The changes to the OH shoulder in the O1s spectra largely follows the same trends indicated in the C1s spectra. To further support the trend of carbonate gain and loss, the in-situ electrical gas sensitivity plot (whereby the basic $CO_2$ adsorption would introduce new charges to the surface of the indium oxide) in **Figure 4d** show that under thermal conditions there is an electrical conductivity increase at a low dose of 0.001-0.002 $CO_2$/Ar flow ratio and subsequently decreases with higher ratio. The lower $CO_2$ gas sensitivity in low doped and high doped $Bi_zIn_{2-z}O_{3-x}(OH)_y$ is also indicated by the smaller increase in conductivity and gradual decrease past the current saturation threshold.

With the above information derived from reactor based investigations of the interaction of $H_2$ and $CO_2$ individually with undoped, low doped and high doped $Bi_zIn_{2-z}O_{3-x}(OH)_y$, let us turn our attention to the XPS spectra of these samples exposed to both gases simultaneously. In this context, the C1s XPS results shown in **Figure 5a-c** reveal the formation of large amounts of surface C-O species around 286 eV, on exposure of $In_2O_{3-x}(OH)_y$ and low doped $Bi_zIn_{2-z}O_{3-x}(OH)_y$ to $H_2+CO_2$ gaseous mixtures under dark/ambient temperature conditions. By contrast, carbonate formation under $H_2+CO_2$ under these same conditions for high doped $Bi_zIn_{2-z}O_{3-x}(OH)_y$ is small. Thermal and photothermal treatment causes these surface C-O species to decay, presumably due to reaction-desorption to RWGS products. Interestingly, the C1s spectra for undoped and low doped is significantly broader than that for under $CO_2$ only thermal conditions. In addition to significant C-O species presence, O-C=O and O=C-OH species are more clearly defined, which indicates that the introduction of $H_2$ enhances the adsorption of $CO_2$ as bicarbonate formation under dark/ambient conditions, however, these species are not thermally or photo-thermally stable. Since there is a minimum temperature of 100°C under photoillumination before any CO products are seen, it can be assumed that the bidentate, tridentate and bicarbonates formed in dark ambient temperature are reacted to form CO under reactor conditions.

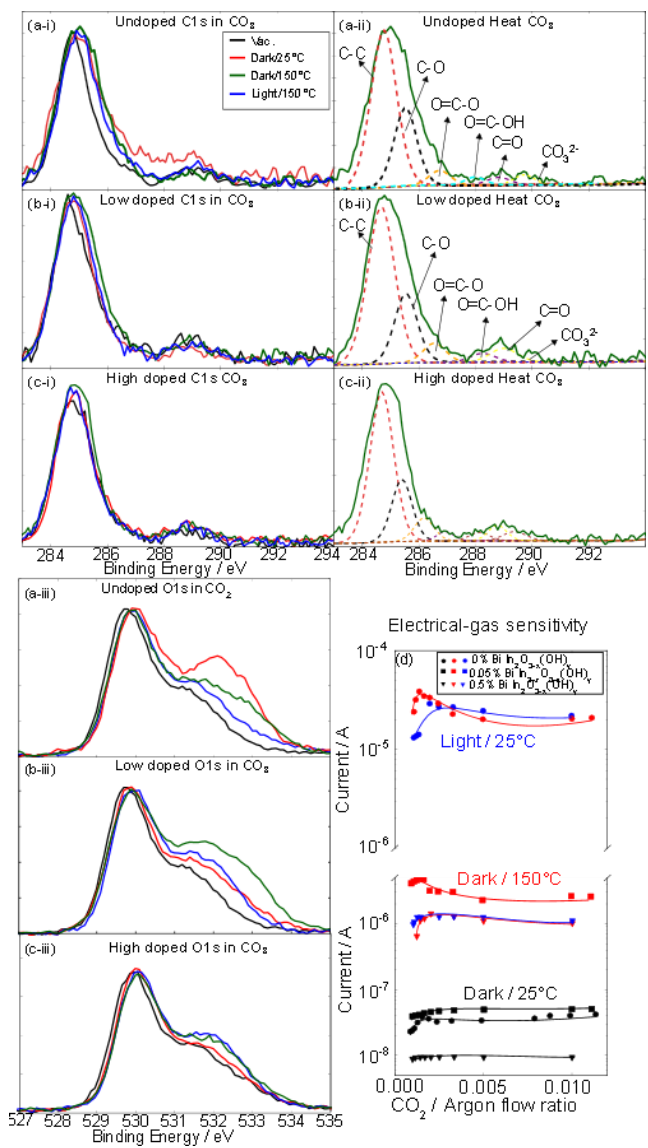

**Figure 4.** C 1s and O 1s XPS of (a-c) of undoped $In_2O_{3-x}OH_y$, for undoped, low doped and high doped under vacuum (black), dark ambient $CO_2$ (red), thermal treatment CO2 at 150°C (green) and photo-illumination, $CO_2$ at 150°C (blue). The deconvoluted C1s spectra of CO2 dark 150C are shown in (ii). (d) Electronic- gas sensitivity measurement of undoped, low doped and high doped $Bi_zIn_{2-z}O_{3-x}OH_y$ under $CO_2$/argon flow ratio. Dark ambient temperature, dark 150°C and photo-illuminated are marked by black, red and blue points respectively. Some trends are omitted for clarity.

The corresponding O 1s and In 3d spectra shown in **Figure 5d-i** for the undoped sample show an even larger expansion of the OH shoulder under photo-thermal condition than that of the $H_2$ only photo-thermal condition, where the OH shoulder has almost the same peak intensity as the $O_L$ peak. Since the $H_2+CO_2$ atmosphere component only introduces carbonate species under dark ambient condition, the expansion of the OH shoulder in the thermal and photothermal condition compared to vacuum state is mostly due to $H_2$ dissociation in $H_2+CO_2$ condition. Interestingly, the sub-peak associated with $O_{vac}$ species also increase with $H_2+CO_2$ from vacuum, and even more so than that of $H_2$ ambient and $H_2$ thermal, which may indicate that $CO_2$ filling of $O_{vac}$ is not prevalent, which supports a Frustrated Lewis Acid-Base Pair type mechanism for the reversed water gas shift reaction. For the undoped sample, the OH shoulder expansion under dark ambient temperature and shrinkage under dark thermal condition is not as prevalent for the same transition for the low doped sample, with less $H^+OH$ specie intensity loss, leading to a fairly broad OH shoulder for the low doped under dark thermal condition. Thus, it is clear that the advantage of low amounts of Bismuth substitutional doping is to enhance $H_2$ dissociation and H adatom thermal stability in a $H_2+CO_2$ atmosphere under photo-thermal conditions. Under $H_2+CO_2$, the low doped $Bi_zIn_{2-z}O_{3-x}(OH)_y$ sample experiences greater changes in the O1s than that under $H_2$ only conditions (**Figure 5d-iii**), which correlates with its higher reactivity to forming CO products. In particular, the OH shoulder expansion leads to an OH shoulder peak intensity greater than that of the $O_L$ peak while In3d peaks under all conditions are normalized equally. In this atmosphere, the large OH shoulder expansion is for low doped $Bi_zIn_{2-z}O_{3-x}(OH)_y$ under photo-thermal condition indicates that the Lewis acidic Bismuth sites exhibit greater photoionization effects than that of the Lewis acidic Indium sites, hence inducing a greater charge polarity between dissociated $H^+$ adatoms on OH groups and oxygen lattice sites, and dissociated $H^-$ adatoms on indium sites.

For the highly doped $Bi_zIn_{2-z}O_{3-x}(OH)_y$ there is only a small expansion in the OH shoulder under photo-thermal condition. While the undoped and low doped In3d peaks exhibit positive binding energy shifts under most reactor conditions, the high doped In3d experiences a clear broadening of the peaks in the positive and negative directions. The In4d peaks also show a lower binding energy tail under thermal and photo-thermal conditions. Furthermore, the Bi4f (**Figure 5j**) spectra exhibits broadening in the dark ambient and photo-thermal conditions. All these deviations for the metallic elements in high doped $Bi_zIn_{2-z}O_{3-x}(OH)_y$ show a reduction to a small concentration of metallic phase, as was shown in the $H_2$ only case. When we plot the generally positive $O_L$ and OH binding energy shifts as a function of the average Pauli electronegativity values with more electronegative bismuth substitution in **Figure 5k**, with decreasing positive increase in binding energy, which again confirms that bismuth substitution decreases Lewis basicity.

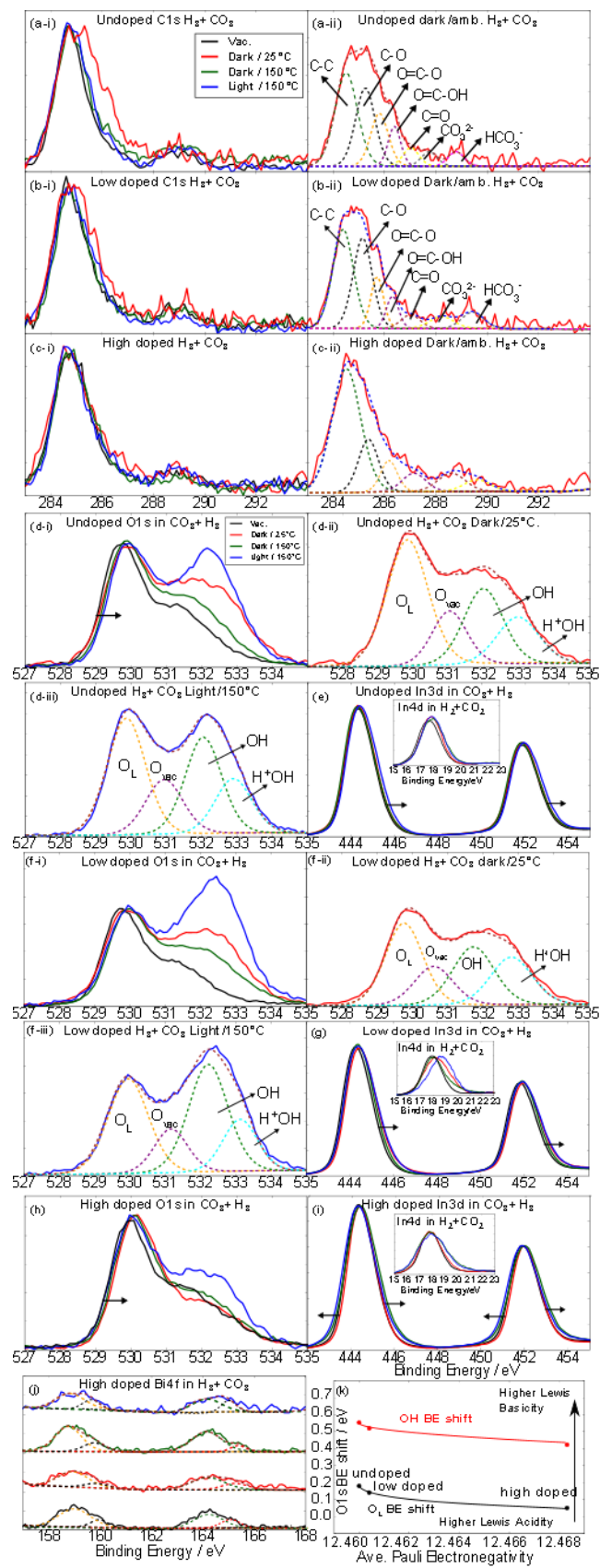

**Figure 5.** C 1s XPS of (top) undoped $In_2O_{3-x}OH_y$, (middle) low doped and (bottom) high doped under vacuum (black), dark ambient $H_2+CO_2$ (red), thermal treatment,

$H_2+CO_2$ at 150°C (green), photo-illumination, $H_2+CO_2$ at 150°C(blue). (a, b, c-ii) shows the deconvoluted C 1s XPS of the $H_2+CO_2$ dark ambient condition. (d-j) O 1s, In 3d and valence band edge XPS of undoped $In_2O_{3-x}OH_y$, under vacuum (black), ambient $H_2+CO_2$ (red), thermal treatment, $H_2+CO_2$ at 150°C (green), photo-illumination and 150°C, $H_2+CO_2$ (blue). The $O_L$ and OH binding energy shifts as a function of the averaged metallic ion Pauli electronegativity is shown.

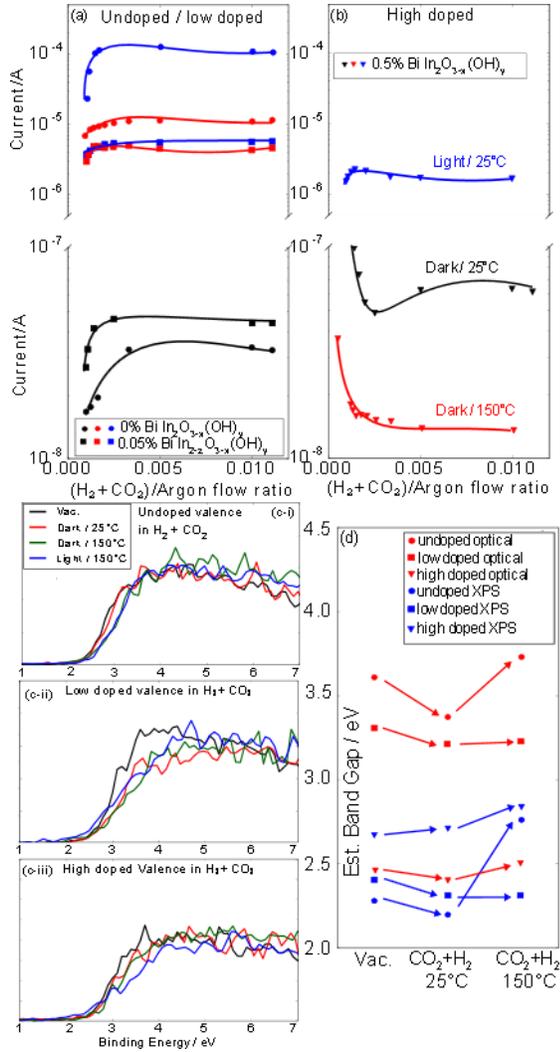

Figure 6. Electronic gas sensitivity measurements for undoped (circle marker) and low doped (square marker) (a) and highly doped sample (triangle marker) (a-b) under a varying set of (1 sccm $H_2$ + 1 sccm $CO_2$) / Argon flow ratio for dark ambient temperature (black), photo-illuminated (blue) and thermal 150°C (red) conditions. (c) Valence band edge of undoped, low doped and high doped in $H_2+CO_2$ atmosphere. (d) UV-Vis band gap measurement in red along with fermi energies determined from valence band edge spectra in blue for undoped (circle), low doped (square) and high doped (triangle) $Bi_zIn_{2-z}O_{3-x}OH_y$ in vacuum, $CO_2+H_2$ ambient temperature and $Co_2+H_2$ mix at 150°C. Trend lines are visual guides.

In **Figure 6**, the valence band edges show significant changes especially under photo-thermal condition, where even more sub-band gap states are present in the low doped $Bi_zIn_{2-z}O_{3-x}(OH)_y$ such that the valence tail reaches up to 1.5eV into the band gap. As support, electronic gas sensitivity measurements in **Figure 6b** show a marked decrease in conductivity for the high doped $Bi_zIn_{2-z}O_{3-x}(OH)_y$ under room and thermal conditions, indicating metallic reduction by $H_2$ reducing gas that increases charge carrier scattering. However, under photo-conditions, the conductivity generally increases, indicating that photo-induced $H_2$ dissociation is contributing electrons to the surface. Additionally, the UV-Vis spectroscopy derived band gap trend also largely follows the valence band edge derived band gap values, which again confirms that the in-situ $H_2+CO_2$ and quasi in-situ $H_2+CO_2$ measurements are mostly equivalent.

Since the typical RWGS reactor testing sequence for $In_2O_{3-x}(OH)_y$ samples have a $H_2$ pressurization and photo-thermal step for one hour before introducing the 1:1 $H_2+CO_2$ pressure mix under photo-thermal conditions in order to characterize CO production using a gas chromatograph, we can chart the energy landscape through the O1s $O_L$ binding energy shifts from vacuum to $H_2$ photo-thermal and subsequently to $H_2+CO_2$ photo-thermal conditions shown in **Figure 7**. Because the binding energy shift of the $O_L$ peak is generally more positive for $H_2$ only than $H_2+CO_2$ mixture, the initial protonation event associated with $H_2$ introduction is stronger than that of the second protonation event of $H_2+CO_2$ introduction. The binding energy decrease from the first protonation event to the second protonation event is smaller for the undoped sample than the low doped $Bi_zIn_{2-z}O_{3-x}(OH)_y$ with -0.02 eV and -0.042 eV respectively. The change in binding energy shifts from $H_2$ to $H_2+CO_2$ for the high doped $Bi_zIn_{2-z}O_{3-x}(OH)_y$ is miniscule at ~ 0.01eV. We note that the aforementioned binding energy shift differences for the low doped sample is a factor of 2.1 that of the undoped sample, which correlates with the CO production rate difference of nearly a factor of 2. Hence the binding energy landscape during the reactor condition transitions likely correlates with the CO rates, analogously with an energy transition over an activation barrier to a lower energy level. The poor CO rate for the high doped $Bi_zIn_{2-z}O_{3-x}(OH)_y$ is thus likely due to $H_2$ to indium reduction and much smaller protonation degree.

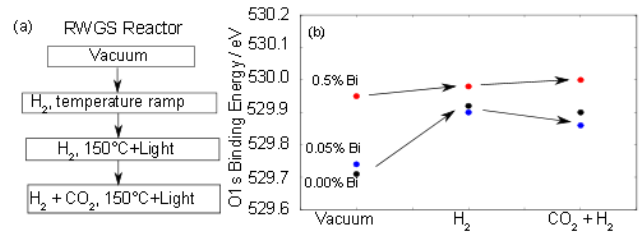

Figure 7. The reactor sequence for the photo-thermal reversed water gas shift reaction is shown in (a). The corresponding binding energy shifts for O 1s oxygen lattice subpeak from vacuum state to $H_2$ photo-thermal 150°C, to $CO_2+H_2$ photo-thermal conditions (b).

**Undoped, low doped, high doped $Bi_zIn_{2-z}O_{3-x}OH_y$ are shown in black, blue and red respectively. The arrows show the first protonation and second protonation events from vacuum to $H_2$ and $H_2$ to $CO_2+H_2$.**

Taken together, the XPS results of the various gas atmospheres show that ambient temperature and dark conditions allow for carbonate and hydroxyl surface formation, with the undoped sample showing greater surface adsorption. However, thermal stability of the generated $H^+OH$ sites by hydrogen dissociation on OH groups is less for the undoped sample than the low doped sample, highlighting one advantage of bismuth substitutional doping, while the photo-thermal condition generates the same amount of OH and $H^+OH$ groups as undoped $In_2O_{3-x}(OH)_y$ at saturated $H_2$ conditions. These results are confirmed by electrical-gas sensitivity measurements, which show that low doped $Bi_zIn_{2-z}O_{3-x}(OH)_y$ have a high degree of $H_2$ adsorption at low concentrations of $H_2$ under photo and thermal conditions and for both $H_2$ and $H_2+CO_2$ atmospheres. This is due to increased overall Lewis acidity of the Bismuth substitution of Indium. The drawback of increased acidity is poorer carbonate adsorption, which is shown by the quasi-operando $CO_2$ XPS measurements. In $CO_2$ only atmosphere, bidentate and tridentate carbonates are more likely formed.

We now discuss the possible permutations of bismuth substitutions in the $In_2O_{3-x}(OH)_y$ lattice and its effect on $H_2$ heterolytic dissociation. As **Figure 8a** shows, an indium site can be attached to (1) the hydroxide group, (2) an oxygen vacancy site and (3) oxide sites. Density Functional Theory calculations of the Bader charge distribution on In, Bi and OH sites in $Bi_zIn_{2-z}O_{3-x}(OH)_y$, for different values of z in the range 0-0.5, revealed that only SFLP configurations *A* and *D* show energetically favorable $H_2$ heterolytic dissociation. A hydroxide bound to Bi1 as shown in configuration *C* is less favorable due to substituted $Bi-In(OH)_3$ having less hygroscopicity than $In(OH)_3$ (see **SI figure S1**), hence there is less likelihood of Bi-OH sites acting as a Lewis base in the $H_2$ heterolysis step. In configuration *B* favored at low values of z, despite the enhanced Lewis acidity of Bi2, heterolysis of $H_2$ could not occur due to the decreased Lewis basicity of In1OH. By contrast, at low values of z in the formula $Bi_zIn_{2-z}O_{3-x}(OH)_y$, where configuration *D* is abundant, the Bi3 site in $Bi_zIn_{2-z}O_{3-x}OH_y$ increases the Lewis basicity of the In1OH hydroxide site and the Lewis acidity of the coordinately unsaturated In2 site, facilitating heterolytic dissociation of $H_2$ and an overall faster RWGS reaction rate compared to $In_2O_{3-x}(OH)_y$. At high values of z, where sites Bi substitution of In2 and In3 sites in configuration *E* abound in $Bi_zIn_{2-z}O_{3-x}(OH)_y$, the Lewis basicity and the Lewis acidity both suffer a decrease in strength which slows the rate of the heterolytic dissociation of $H_2$. In **Figure 8b**, a simplified energy band diagram shows the possible levels of Lewis acidic $In^{3+}$ and $Bi^{3+}$ sites, and Lewis base sites of OH lying below the conduction band and above the valence band respectively. As was shown in the valence band edge, $H^+OH$ specie growth is correlated with having a longer valence band tail, where the OH groups act as hole acceptor states.

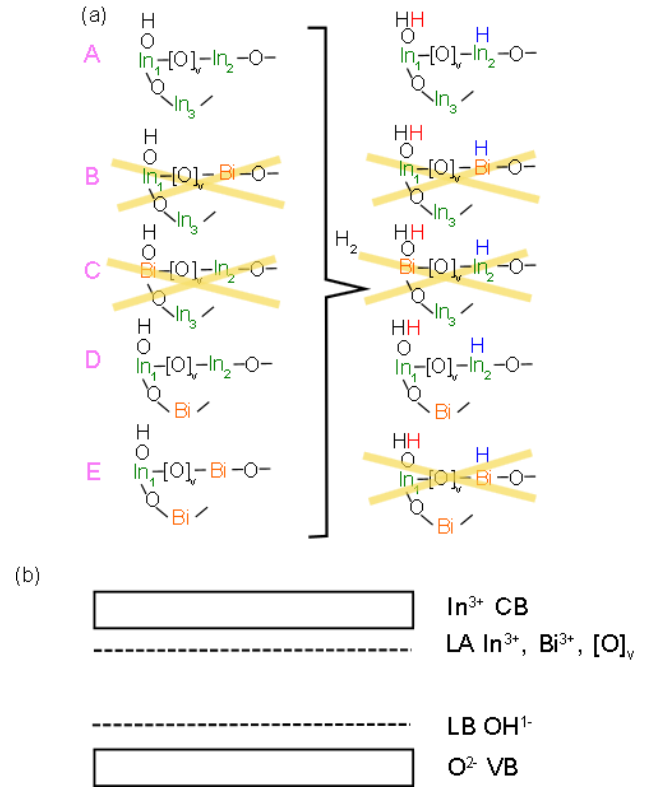

**Figure 8.** (a) Illustration of Surface Frustrated Lewis Pairs in $Bi_zIn_{2-z}O_{3-x}OH_y$ for $0 < z < 0.5$: the schematics are simplified planar arrangements on the surface (some elements omitted for clarity) where the lines between elements represent the bonds roughly planar to the surface while the absent lines especially linking OH represent terminal groups roughly perpendicular to the surface. The various configurations: (A) depicting Lewis acid In1 and Lewis base In2 surface sites and a proximal sub-surface site In3 in the parent material $In_{2-z}O_{3-x}OH_y$, with substitution (B) In2 by Bi2, (C) In1 by Bi1, (D) In3 by Bi3, and (E) In2 and In3 by Bi. The heterolysis (right) of $H_2$ on each of these sites. The yellow crosses represent species that are deemed unreactive (B,E) or do not exist (C), as explained in the text. Red H adatoms represent $H^+$ and blue H adatoms represent $H^-$ species. (b) Electronic band diagram of $Bi_zIn_{2-z}O_{3-x}OH_y$ depicting the mid-gap hydroxide, coordinately unsaturated indium and bismuth, oxygen-vacancy defect sites, straddled below by the oxide valence band and above by the indium conduction band

We now discuss photo-effects on the $Bi_zIn_{2-z}O_{3-x}(OH)_y$ system. From in-situ photo-EPR measurements on $In_2O_{3-x}OH_y$ it was shown[17] that new paramagnetic centres not associated with that of singly ionized oxygen vacancies are generated under blue light but are not seen once the blue light is turned off. These new centres are associated with valence band holes from photoexcitation and can thus be associated with the elementary photo-ionization transitions: $[O]^._{vac} + h\nu \rightarrow [O]^{..}_{vac} + h^+_{VB}$; $In^{3+} + h\nu \rightarrow In^{2+} + h^+_{VB}$; and in this doped system we also propose the following photo-transition $Bi^{3+} + h\nu \rightarrow Bi^{5+} + 2h^+_{VB}$. We thus suggest that under $H_2$ atmosphere an indium-hydroxide or indium-oxide site adjacent to an oxygen vacancy site accepts a valence hole which stabilizes the dissociated $H^+OH$ and $H^+O$ sites respectively under high temperatures. This would explain the increase in OH and HOH XPS peaks under photo-thermal but not thermal conditions. Photoionization of $Bi^{3+}$ would yield 2 valence holes and thus stabilize both types of sites, as well as present $Bi^{5+}$ states which are seen in the XPS spectra. Since the proposed SFLP reaction supposedly proceeds by the following reaction: $HO-In-[O]-In + H_2 + CO_2 \rightarrow HOH^+-In-[O]-In-H^- + CO_2 \rightarrow HOH^+OOC-H^- \rightarrow CO + H_2O + HO-In-[O]-In$, the reaction would remove $H^+OH$ sites while retaining $OH^+$ sites, hence the OH peak is larger than that of the HOH peak, as confirmed by the O1s spectra under $H_2+CO_2$ photothermal condition.

**Conclusion**

To obtain a detailed atomic level understanding of gas-phase, light-assisted heterogeneously catalyzed hydrogenation reactions of $CO_2$ to chemicals and fuels, such as CO, $CH_4$, $CH_3OH$ and $C_nH_{2n+2}$, requires operando analytical techniques. These include microscopy, diffraction, spectroscopy and electrical probes of the structure and reactivity, vibrational, optical and electrical properties of adsorbed species on the surface of photocatalysts. The goal is to reveal the identity of adsorbed species and their behavior under reaction conditions.

In this article, for the first time quasi in-situ XPS and other in-situ measurements interrogates the photochemical RWGS reaction under conditions close to those occurring in a photocatalytic reactor. The results provide new information on the reactions of nanostructured $Bi_zIn_{2-z}O_{3-x}OH_y$ with $H_2$, $CO_2$ and $H_2+CO_2$ under thermochemical and photochemical conditions. The shifts in energy and changes in intensity of the C 1s, O1s, In 3d, Bi 4f ionization potentials provide information on the dissociation of $H_2$ on surface Lewis base hydroxide and Lewis acid coordinately unsaturated metal sites, and the reaction of $CO_2$ and $H_2+CO_2$ mixtures with these surface sites. With an examination of the binding energy shifts we are able to chart an energy landscape corresponding to the RWGS reactor operation. The advantage of isolating the various reactor conditions allow examination of the important roles of photo and thermal effects, whereas in most catalytic studies, such effects are generalized as activation energy barrier studies. In all, this study encourages detailed photo-thermal catalysis chemical studies that lead to better performing catalysts.

## ASSOCIATED CONTENT

### Catalyst Preparation

To prepare the z = 0.5% $Bi_zIn_{2-z}O_{3-x}OH_y$ : solution A was prepared by adding 1.8 mmol of $In(NO_3)_3$ and 6 ml ethanol and 1ml $Bi(NO_3)_3$ solution and 1 ml $H_2O$. Solution B was prepared by adding 2 ml $NH_3.H_2O$ with 6 ml ethanol solution. Add solution A to solution B, stir for 1 minute and heat in an oil bath set to 80°C for 10 minutes. After cooling, centrifuge and wash with deionized water for 5 rounds. Dry in vacuum oven and calcine the product at 250°C for 6 hours.

### Experimental Section

The reactors were vacuum pumped till 2.5×10⁻² mBarr, then the gases were pumped in till a pressure of 2atm. For the mix atmosphere, ~1 atm of $H_2$ was introduced before introduction of additional 1 atm of $CO_2$. The reactors were separated into dark room temperature or photo-illuminated with a 150W white xenon/mercury lamp or thermally treated at 150°C in dark, for 2 hours. The reactors were carried to the glovebox attached XPS and opened within the glovebox.

### Characterizations

The UV-Vis instrument was a Perkin Elmer Lambda 1100, the diffuse reflectance spectra were taken in a dark room using a reactor with temperature control, where the glass window spectra was auto-zeroed. The XPS instrument was a Thermo Scientific Theta Probe XPS system.

## AUTHOR INFORMATION

### Corresponding Author

* Kherani@ecf.utoronto.ca
*Prof Nazir P. Kherani*

### Author Contributions

Joel Y.Y Loh performed XPS data analysis, Electrical, UV-Vis spectroscopy experiments and their design. Joel Y.Y Loh also prepared the current manuscript. Yufeng Ye modified the Tauc absorption code for use in this study and provided edits and feedback to the manuscript. Nazir P. Kherani is the principal investigator.

### Funding Sources

The authors gratefully acknowledge the support of the Natural Science and Research Council of Canada.